\documentstyle{elsart}
\begin{document}

\input epsf.sty

\begin{frontmatter}
\title{Collapse of a polymer in two dimensions}

\author[Delft]{H.W.J. Bl\"ote}
\author[Canberra]{M.T. Batchelor} and
\author[Amstel]{B. Nienhuis}
\address[Delft]{Faculteit Technische Natuurkunde,
Technische Universiteit Delft, Postbus 5046, 2600 GA Delft, 
The Netherlands}
\address[Canberra]{Department of Mathematics, 
School of Mathematical Sciences,
Australian National University, Canberra ACT 0200, Australia}
\address[Amstel]{Instituut voor Theoretische Fysica,
Universiteit van Amsterdam, Valckenierstraat 65, 1018 XE Amsterdam, 
The Netherlands}

\begin{abstract}
We numerically investigate the influence of self-attraction 
on the critical behaviour of a polymer in two dimensions, by means of
an analysis of finite-size results of transfer-matrix calculations.
The transfer matrix is constructed on the basis of the O($n$) loop model
in the limit $n \rightarrow 0$. It yields finite-size results for 
the magnetic correlation length of systems with a cylindrical geometry.
A comparison with the predictions of finite-size scaling enables
us to obtain information about the phase diagram as a function of the
chemical potential of the loop segments and the strength of the 
attractive potential.
Results for the magnetic scaling dimension can be interpreted
in terms of known universality classes. In particular,
when the attractive potential is increased, we observe the crossover 
between polymer critical behaviour of the self-avoiding walk type to 
behaviour described earlier for the theta point.
\end{abstract}
\end{frontmatter}

\section{Introduction}
A useful formulation of the polymer problem can be given in terms of the 
O($n$) loop model on a lattice, in the limit $n \rightarrow 0$.
In this formulation, loops have a weight $n$ so that the number of 
loops is minimized as $n \rightarrow 0$. In the simplest case, only
one other parameter plays a role: a weight factor 
$a$ may be assigned to each loop segment, in order to control the density
of the loop configurations. A loop segment is the part of a loop that 
covers precisely one lattice edge. Loops do not intersect.
For small $a$ the vacuum state -- without any loops -- is stable; 
in contrast, 
for large $a$, most of the lattice edges are covered by a loop. 
The critical point separating the vacuum state and the dense phase
has already been explored in detail; it is found to
belong to the same universality class as the SAW (self-avoiding walk).  

Exact results for the O($n$) loop model on the honeycomb lattice have 
been obtained \cite{NienhuisOn,Baxter,Batchelor-B,Suzuki} that reveal 
two branches of critical points (two points for each $n$ in the
range between 0 and 2). One of these branches is interpreted
as a critical point separating the disordered and long-range
phases; the other branch is supposed to describe the ordered phase.
This branch has algebraically decaying correlations for general $n$;
in this sense it still qualifies as a critical branch.

This picture applies to the case where the loop segments do not interact.
A natural extension of the noninteracting loop model is to adjust the
vertex weights in a way representing attractive forces between loop
segments. Such interactions, if they are of a sufficient strength,
will influence the character of the phase transition between the vacuum
and the dense phase. In the case of strong attractive forces, the critical 
state, which is relatively dilute, becomes unstable, so that the phase
transition becomes first-order.

The higher order critical point, separating the first-order transition 
from the continuous one, describes a polymer on the verge of
collapse, and is called the theta point. 
An exactly known critical point of an $n=0$ loop model with vacancies
on the honeycomb lattice, described by Duplantier and Saleur \cite{DS}, 
belongs to the universality class of this theta transition. 

The formulation of the square O($n$) loop model given in 
Ref.~\cite{BNjpa}
contains three adjustable parameters. One of these applies to the weight
of vertices visited twice by a loop. There, both loop parts make 
$90^\circ$ bends such that they collide but do not intersect. Thus, 
by varying this vertex weight one can tune the attractive forces.

For this three-parameter model, there are several branches of exactly 
known critical points \cite{BNjpa,BNW,N90,WBN}. 
Two of these have the same universal
properties as those mentioned above for the honeycomb lattice.
Two other branches are associated with a combination of Ising-like
and O($n$) critical behaviour. The fifth branch, which was dubbed
`branch 0', is related with the dense phase of the O($n+1$) model
\cite{BNjpa} and contains an integrable critical point \cite{BNjpa,B93}
similar to the above-mentioned theta point \cite{DS}.
In particular, the O($n=0$) point has been shown to be
equivalent to interacting walks on the Manhattan lattice at
the theta point \cite{PO}, with a set of exponents in agreement
with those proposed for the theta transition \cite{DS},
if one associates the magnetic dimension $x_m=\eta/2=0$ \cite{DS} 
with the critical dimension $x_{{\rm int},1}$ which was discussed in
Ref.~\cite{BNjpa}.

Thus, the subset of the parameter space contains a considerable 
variety of critical points for which exact information is 
available \cite{BNjpa,BNW,N90,WBN,B93,PO,bosy}.
Nevertheless, it
covers only a small part of the parameter space. Therefore, it is of 
interest to explore the critical surface by numerical means. In particular,
here we investigate the influence of attractive forces between the loop
segments in the square O($n$) loop model, in the hope to observe the 
crossover between the normal O($n=0$) or SAW-like critical point and the
proposed theta point.

In Section 2 we briefly introduce the O($n$) loop model, and the 
transfer matrix used for the numerical calculations. Section 3 
contains an analysis and a discussion of the numerical work. For this 
purpose, it also summarizes the finite-size scaling formulas and an 
element of the theory of conformal invariance used in the analysis.

\section{The model and its transfer matrix}

The loop representation of the O($n$) model yields an expression
for the partition function in terms of a sum over all graphs 
consisting of closed loops, covering a subset of
the lattice edges \cite{dmns,NienhuisOn}. The simplest case, with
only two parameters, the loop weight $n$ and the weight factor $a$ per 
bond, has been investigated for the O($n$) loop model on the honeycomb 
lattice. In addition to exact results, conclusions based on a numerical 
investigation are available. Finite-size scaling of transfer-matrix 
results can be applied in parts of the parameter space where no
exact information is known. These results \cite{BNphy} have confirmed 
the interpretation of the two critical branches as mentioned above. 

In this work, we apply a similar method to the square lattice. The square
O($n$) loop model is now formulated in terms of loop weights
$n$, bond weights $a$, and an additional weight factor $p$ associated
with each straight vertex, and an additional weight factor $q$
associated with each `collision'. Thus, the vertex weights are
$w_0=1$ for an empty vertex (not visited by a loop), $w_1=a$ for
a vertex visited once by a loop making a $90^\circ$ bend, $w_2=ap$ 
for a vertex visited once by a straight loop part, and $w_3=a^2q$
for a vertex visited twice (see Fig. 1). Using these weights, the 
partition function becomes
\begin{equation}
 Z  =  \sum_{\cal G} w_1^{N_1} w_2^{N_2}  w_3^{N_3} n^{N_l}  ,
\label{loopZ}
\end{equation}
where the sum is over all loop configurations ${\cal G}$. 
The graph ${\cal G}$ consists of $N_l$ nonintersecting loops,
with $N_i$ ($i=1,2,3$) the total 
number of vertices of the type indicated by index $i$. 
\begin{figure}[h]
\vskip 5mm
\centerline{
\epsfxsize=9cm
\epsfbox{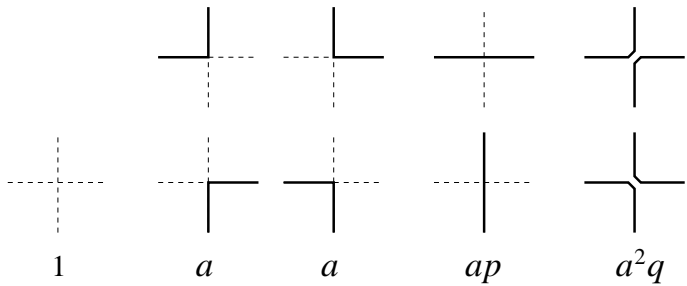}}
\vskip 5mm
\caption{Vertex weights for the O($n$) loop model on the square 
lattice. In the present work we choose $p=0$, so that $180^\circ$ 
vertices are excluded.
}
\end{figure}

A transfer matrix can be constructed for this loop model; a detailed
description is given in Ref.~\cite{BNjpa}. The transfer matrix can be 
seen as an operation that builds up a cylinder on which the lattice 
O($n$) model is wrapped, by adding one circular row, and thus increasing
the length of the cylinder by one unit. The presence of the open end of
the cylinder allows non-empty bond configurations even for $n=0$, 
where closed loops are actually excluded.  The transfer matrix indices
are a numerical coding of `connectivities': the way in which the 
dangling bonds, or loop segments, are connected at the end of the
cylinder. The allowed connectivities are `well nested': no four occupied 
edges can be crosswise connected. This property is a consequence of the 
absence of intersections, and greatly restricts the number of 
connectivities. A sparse-matrix decomposition \cite{BNjpa} allowed us to 
obtain transfer-matrix eigenvalues for systems up to linear size 
$L=12$ using only modest computational resources. 

These transfer matrix eigenvalues are
meaningful because they determine the free energy and the length
scales determining the decay of the correlation functions along
the cylinder. 
In the general case, the free energy per unit of area $f$ of a model 
on an infinitely long cylinder with finite size $L$ is determined by
\begin{equation}
 f(L) = L^{-1}  \ln \Lambda^{(0)}_L ,
\label{flambda}
\end{equation}
where $\Lambda^{(0)}_L$ is the largest eigenvalue of the transfer
matrix. In the paramagnetic phase of the O($n=0$) loop model, as well as
on the critical line separating the ordered phase, $\Lambda^{(0)}_L=1$.
The corresponding eigenvector is dominated by the vacuum.
This eigenvalue 1 persists in the ordered phase, but there it is no longer
the largest eigenvalue. For reasons of continuity, we maintain the
notation $\Lambda^{(0)}_L=1$.

For practical reasons, the set of connectivities is
split into two disjoint subsets: the even subset, where all the
dangling bonds are connected pairwise; and the odd subset where,
in addition, a single, unpaired dangling bond occurs. The 
vacuum-dominated leading eigenvector naturally occurs in the even
sector. The odd sector contains a line of covered bonds, running 
along the length direction of the cylinder.
Such a line is not a part of a closed loop, and does not occur
in the graphs ${\cal G}$ of Eq.~(\ref{loopZ}). Therefore, the odd
sector does not contribute to the partition sum.

However, the odd connectivities are important for the calculation of
the magnetic correlation length of the O($n$) model. The graphs containing,
in addition with closed loops, a single line of covered bonds connecting
two points are precisely those describing the magnetic correlation
function between O($n$) spins located on those points. It is expressed
as $Z'/Z$, where $Z$ is given by Eq.~(\ref{loopZ}), and $Z'$ by the same
equation but with ${\cal G}$ replaced by ${\cal G'}$ representing all 
graphs consisting of closed loops plus a single line connecting the two 
spins. 

Thus, the inverse magnetic correlation length $\xi^{-1}(L)$ can be
expressed in terms of $\Lambda^{(0)}_L$ and $\Lambda^{(1)}_L$ which
denotes the largest eigenvalue in the odd sector:
\begin{equation}
 \xi^{-1}(L) = \ln(\Lambda^{(0)}_L/\Lambda^{(1)}_L) .
\label{xilambda}
\end{equation}
A convenient quantity in the finite-size analysis is the `scaled
gap' $x(u,L)= L/[2\pi \xi(u,L)]$ whose arguments include, besides the 
finite size $L$, also a temper\\ature-like parameter $u$. For instance we
may choose $u=a-a_c$ where $a_c$ denotes the critical value of the 
bond weight $a$. But $u$ may also be chosen to parametrize the
distance to fixed points located on the critical surface.
In the vicinity of a renormalization fixed point, finite-size
scaling leads to the equation
\begin{equation}
 x(u,L) = x_h + 
\frac{1}{2\pi} L^{y_u} u [{\rm d}\xi^{-1}(u,1)/{\rm d}u]_{u=0} + \ldots ,
\label{X-scaling}
\end{equation}
where $x_h$ is the magnetic scaling dimension \cite{Cardy-xi} of
the corresponding fixed point, and $y_u$ the renormalization exponent
associated with $u$, governing the flow to or away from the fixed point.
Thus, if $y_u<0$ ($>0$), $x(u,L)$ converges to (diverges from) $x_h$ 
with increasing $L$. This allows us to analyse the
behaviour of the finite-size results $x(u,L)$ in the light of
the phase diagram. 

\section{Results}

Before starting the actual transfer-matrix calculations at $n=0$, we 
summarize the role of the three parameters. First, the bond weight $a$
(or $w_1$) is adjusted in order to find the critical point. Expressing
the scaled gap in $a$, ignoring the irrelevant scaling fields, and
using finite-size scaling of the correlation length,
the critical point $a_c$ is determined by numerical data for two 
subsequent even finite sizes $L$ and $L+2$:
\begin{equation}
 x(a_c,L) = x(a_c,L+2) .
\label{ac}
\end{equation}
Only even sizes are used because, in general, the odd systems display
different finite-size amplitudes \cite{BNjpa}. Corrections to scaling
introduce deviations from Eq.~(\ref{ac}) so that extrapolation of the
finite-size estimates of $a_c$ was performed. For details, see e.g.
Ref.~\cite{GBB}.


The parameter $p$ is here important with regard to the
Ising-like degrees of freedom that play a role in the square loop
model \cite{BNjpa}. In order to clarify this point, we introduce an 
Ising spin on each elementary face, such that two neighboring
spins are equal only when a loop passes between them. Clearly when 
$p=0$ a loop is adjacent to spins of one sign only, signalling a
broken Ising symmetry. This means that an Ising-like ordering 
transition occurs when $p$ is lowered. 

In the present work, we wish to focus on the collapsing polymer
problem, while avoiding the interfering effects associated with 
Ising-like ordering.
A possible way to circumvent these effects is to exclude type
2 vertices by putting $p=w_2=0$, so that the Ising degrees of freedom
are already frozen
even in the relatively dilute SAW-like critical state. An 
increasing polymer density does not further affect this Ising ordering.

Thus, we scanned the parameter space using $w_1=a$, $w_2=0$, and
$w_3=q a^2$ where the parameter $q$ governs the attraction between 
the loop segments. This choice of parameters includes the integrable 
theta point mentioned above, for which $a=0.5$, $q=2$.

Using finite size parameters $L=2$, 4, 6, 8 and 10, Eq.~(\ref{ac})
was solved for $q=0$, 1.0, 1.5, 1.8, 1.9, 2.0, 2.1, 2.2, 2.5, 3.0
and 4.0. The extrapolated critical points are shown in Table 1.

\begin{table}                                                              
\caption{
Numerical results for the critical value of the bond weight $a$ 
for O($n=0$) loop models with different values of the loop-loop 
interaction parameter $q$ (see column 1). For $q\leq1.5$ we assumed 
a correction term proportional to $L^{-9/4}$, for $q>1.5$ one
proportional to $L^{-1}$. The rightmost column shows the estimated
value for the magnetic  dimension $x_h$ where extrapolation was
possible. The extrapolations for $q=2$ are in a good agreement with
the exact values $a_c=1/2$ and $x_h=1/4$.
\vskip 5mm
}
\centerline{
\begin{tabular}{||c|lc|lc||}  
\hline                                                                    
 $q$    &    $a_c$ &    & $x_h$  &     \\
\hline                                                                      
   0.0  &  0.63860 &(5) & 0.1045 & (5) \\  
   1.0  &  0.5769  &(1) & 0.104  & (1) \\  
   1.5  &  0.5399  &(1) & 0.11   & (1) \\  
   1.8  &  0.5165  &(1) &        &     \\  
   1.9  &  0.5084  &(1) &        &     \\  
   2.0  &  0.5001  &(1) & 0.2500 & (2) \\  
   2.1  &  0.4917  &(1) &        &     \\  
   2.2  &  0.4833  &(2) &        &     \\  
   2.5  &  0.4575  &(5) &        &     \\  
   3.0  &  0.421   &(1) &        &     \\  
   4.0  &  0.368   &(1) &        &     \\  
\hline                                                                      
\end{tabular}}
\end{table}

The value $a_c=0.63860(5)$ at $q=0$ gives the connective 
constant $1/a_c\simeq1.565\ldots$ for SAWs with a $90^\circ$ turn at
every step, in good agreement with the estimate 1.5657(19) \cite{G}. 
The scaled gaps at the intersections are plotted versus
the system size $L$ in Fig. 2. These data are to be compared to
Eq.~(\ref{X-scaling}) where we may interpret $t$ as a field parametrizing
the critical line; the leading temperature field vanishes in effect
for the solution of Eq.~(\ref{ac}). The field along the 
critical line is expected to be irrelevant in the vicinity of the 
SAW-like fixed point; at a higher critical point, it is expected 
to be relevant. Indeed, for $q<2$ we observe, for increasing $L$,
a converging trend towards the
exactly known value $x_h=5/48=0.104166\ldots$ for the SAW model.
This convergence reflects the stability of the SAW-like fixed point.
For $q=2$ the finite-size data converge well to $x_h=1/4$, in 
agreement with the known value at the integrable point.

\begin{figure}[h]
\centerline{
\epsfbox{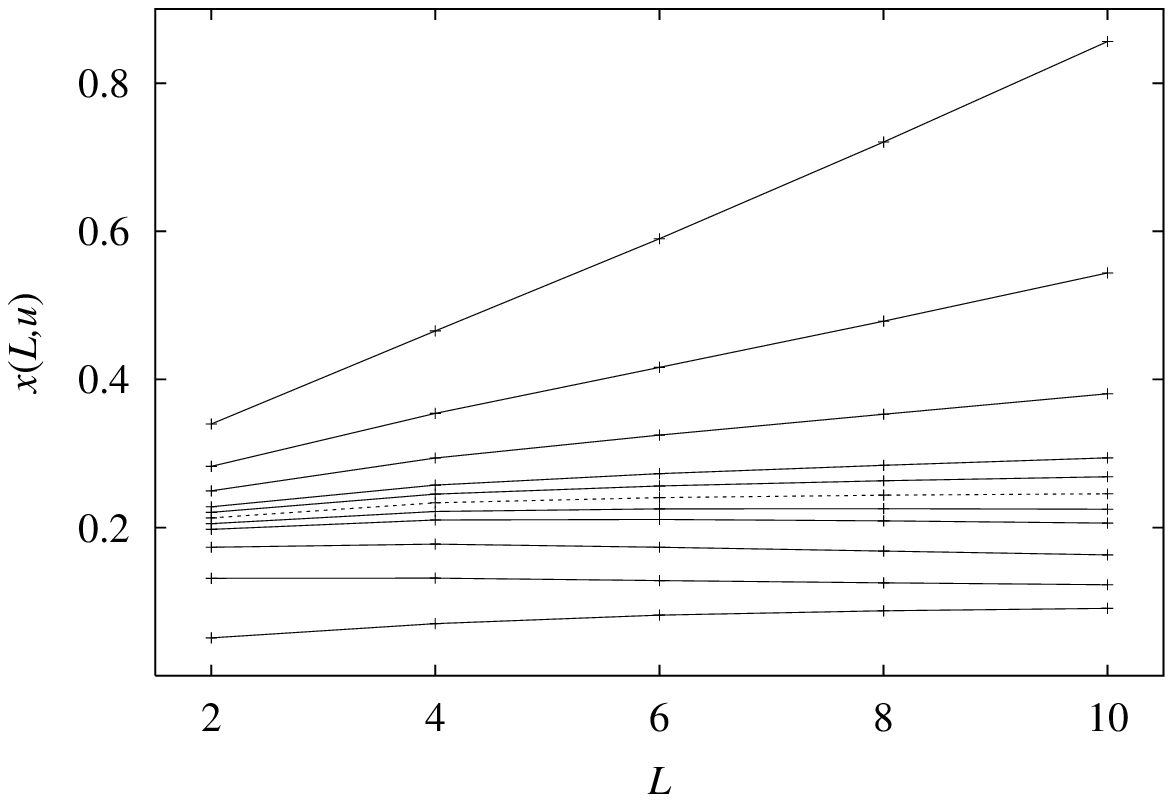}}
\caption{Finite-size dependence of the scaled gap of O($n$) models.
These data are taken at values of the bond weight $a$ solving
the scaling equation for the correlation length for two subsequent
system sizes. The lines connecting the data points are for visual
aid only. Starting from below, the lines apply to $q=0$, 1.0, 1.5, 
1.8, 1.9, 2.0 (dashed line), 2.1, 2.2, 2.5, 3.0 and 4.0 respectively.
}
\end{figure}
For $q>2$ the phase transition becomes discontinuous as a
function of the temperature-like parameter $a$, as is revealed
by an intersection of the two largest eigenvalues $\Lambda_0=1$ and
$\Lambda_2$ of the transfer matrix in the even sector. For the
interpretation of the results shown in Fig. 2 it is important to 
note that we used $\Lambda_0=1$ in Eq.~(\ref{xilambda}), even where 
other eigenvalues exceed 1; this occurs for $q>2$ even at
the line of phase transitions. Using this
analytic continuation we avoid irregularities caused by the
intersections mentioned. The increasing trend of the data in Fig. 2
for $q>2$ reflects the instability of the theta-like fixed point.

However, the lines connecting the finite-size results in Fig. 2
are running almost horizontally for $q\approx 2$; the finite-size 
dependence of the scaled gap is rather weak.
This corresponds with a small value of the exponent $y_u$ in 
Eq.~(\ref{X-scaling}) when applied to the theta point. Indeed, from 
Coulomb gas arguments \cite{DS,n} one expects $y_u=3/4$.

Also in the vicinity of the stable SAW-like fixed point, the finite-size
dependence of the data shown in Fig. 2 is rather weak. This
stands in a strong contrast with the rapid convergence observed 
\cite{hbu} for O($n=0$) loop models with $p=1$, which can be 
interpreted \cite{GBB} in terms of an irrelevant exponent $y_i=-2$.
The present slow trends near the SAW-like fixed points may be
explained by an irrelevant exponent $-11/12$, in analogy
with the case of `trails' where loops are allowed to intersect
\cite{GBB}. In the Coulomb gas representation, loop intersections 
and collisions correspond with the same diagrams.

We have performed extrapolations on the basis of the numerical 
solutions of Eq.~(\ref{ac}). According to finite-size scaling, 
the solutions $a(q,L)$ obtained from finite sizes $L$ and $L-2$
behave as $a(q,L)=a(q)+b L^{y_u-y_t} +\cdots$ where $y_t$ is the
leading temperature exponent; $y_t=4/3$ at the SAW-like fixed
point, and  $y_t=7/4$ at the theta point. For the second
temperature-like exponent we may expect $y_u=-11/12$ and $y_u=3/4$ 
respectively. On the basis of our limited range of finite sizes, it
was not possible to well determine $y_u$ independently 
for all $q$. Only in the case $q=0$ we could confirm that $y_u-y_t
\approx -2.2$ as expected. For $q=2.1$ and 2.2 we find results for
$y_u-y_t$ consistent with $-1$ but the data show that further corrections
are important. Thus we assumed $y_u-y_t=-9/4$ for $q\leq1.5$ and 
$y_u-y_t=-1$ for $q > 1.5$, and extrapolated the finite-size estimates 
of the critical points accordingly. The results are shown in Table 1.
More details of the fitting procedure can be found in Ref.~\cite{GBB}.

Since the solutions of Eq.~(\ref{ac}) converge to the critical point,
the scaled gaps converge to the corresponding magnetic scaling
dimension. For those cases where the finite-size convergence was
sufficiently fast, extrapolated results are included in Table 1. 
The results for $q=0$, 1 and 1.5 are in a good agreement with the
expected O($n=0$) magnetic dimension $x_h=5/48= 0.104166\ldots$
Thus, the fact that the Ising degree of freedom is frozen does not 
seem to alter the universal properties.  
For $q=2$ our result for the magnetic 
dimension agrees well with $x_h=1/4$ as found earlier for this integrable
point of the square O($n=0$) model \cite{BNjpa,B93}. For larger values of
$q$ the finite-size data assume a monotonically increasing trend with
$L$, in accordance with earlier observations at first-order 
transitions, for instance that of the Potts model with a large number of
states \cite{BNig}.

The phase diagram of the model, summarizing our findings, is shown in 
Fig. 3. The vacuum state occurs near the origin, and a dense phase at 
large values of $w_1$ and $w_3$. These phases are separated by a line of 
phase transitions of which the upper part is first order, and the right
hand part is critical, with the theta point separating both parts.

\begin{figure}[h]
\centerline{
\epsfbox{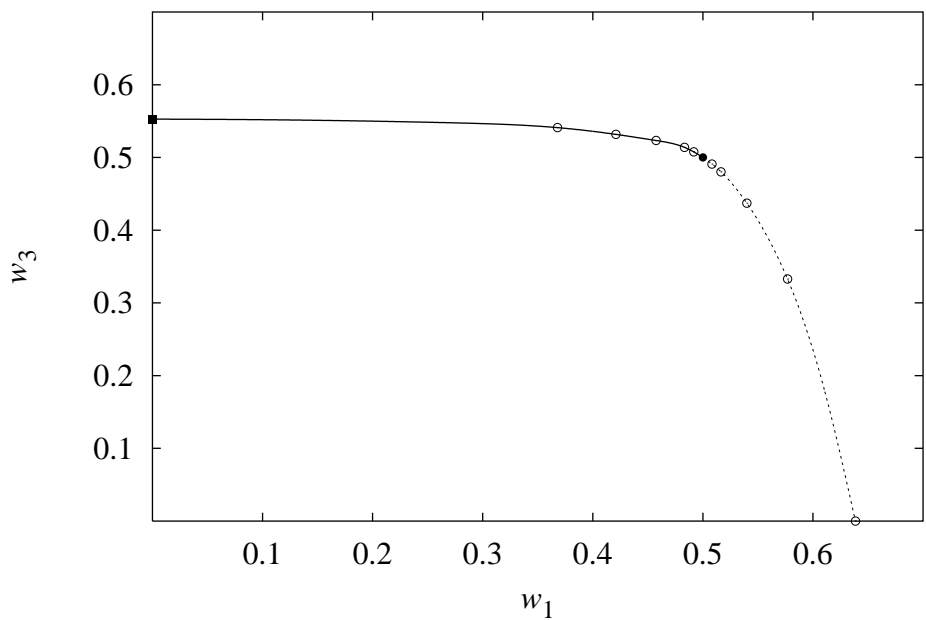}}
\caption{Phase diagram of the O($n=0$) model for $w_2=0$.  
The line of phase transitions has a first-order part (solid line)
and a continuous part (broken line). This line is an interpolation 
between the numerical results (open circles), the exactly
known theta point at $w_1=1/2$ (black circle) and the exactly known 
point at $w_1=0$ (black square).
}
\end{figure}

In conclusion, our data confirm that the introduction of sufficiently
strong attractive forces in the critical O($n=0$) model
leads to the collapse of a polymer in two
dimensions. The critical exponents at the point of collapse are in
agreement with the predictions of Duplantier and Saleur \cite{DS}
provided we avoid the onset of Ising-like ordering which is known to
lead to a different universality class \cite{BNjpa}.

\ack

It is a pleasure to express our appreciation for the privilege of
many stimulating and valuable discussions with Prof. J.M.J. van Leeuwen.
Furthermore we express our deep gratitude for his indefatigable efforts
to provide the right conditions for our research efforts, of which, 
among other things, our present collaboration is a result.

This work has been supported by the 
``Stichting voor Fundamenteel Onderzoek der Materie (FOM)'', which
is financially supported by the ``Nederlandse Organisatie voor 
Wetenschappelijk Onderzoek (NWO)'', and the Australian Research Council.

\end{document}